\documentstyle{article}

\title{\bf
Population Monte Carlo algorithms}

\author{
{\Large Yukito Iba} \\
The Institute of Statistical Mathematics\\
4-6-7 Minami-Azabu, Minato-ku, Tokyo, 1068569, Japan \\
iba@ism.ac.jp}

\date{}

\begin{document}
\sloppy
\maketitle

\vspace{3cm}

\begin{abstract}
We give a {\it cross-disciplinary}
survey on ``population'' Monte Carlo algorithms. 
In these algorithms, a set 
of ``walkers'' or ``particles'' is used
as a representation of a high-dimensional vector.
The computation is carried out by a random walk
and split/deletion of these objects.
The algorithms are developed 
in various fields in physics and statistical sciences
and called by lots of different terms --
``quantum Monte Carlo'', 
``transfer-matrix Monte Carlo'', ``Monte Carlo filter
(particle filter)'',``sequential Monte Carlo'' and ``PERM''
{\it etc.} Here we discuss them in a coherent
framework. We also touch on related algorithms --
genetic algorithms and annealed importance sampling.\\ 
\end{abstract}

\vspace{1cm}

\begin{quote}
{\small {\bf Keywords:}
quantum Monte Carlo, transfer-matrix Monte Carlo,
Monte Carlo filter, sequential Monte Carlo, 
pruned-enriched Rosenbluth method, annealed importance 
sampling, genetic algorithm
}
\end{quote}

\newpage

\section{Introduction}

In this paper, we give a {\it cross-disciplinary\/}
survey on ``population'' Monte Carlo algorithms.
These algorithms, which are developed 
in various fields, have a common 
structure: 
A set of ``walkers'' or ``particles''
is used for the representation of a high-dimensional vector
and the computation is carried out by a random walk in the state space
and split/deletion of these objects. 
These algorithms do {\it not\/} belong to 
the class of Markov chain Monte Carlo (MCMC, dynamical Monte Carlo),
although they share common features and applications.

Monte Carlo filter~\cite{K96,K98} 
(or sequential Monte Carlo~\cite{LC98,book,Crisan00})
algorithm is a topic 
of recent interest in the field of 
statistical information processing and
now being popular in related fields, 
{\it e.g.,} robot vision~\cite{I98}. 
It is an example of population
Monte Carlo algorithms defined here.
It is, however, {\it not\/}
the only example. 
Population Monte Carlo algorithms
have proved to be useful
tools in a number of fields --- 
quantum physics, polymer science, statistical physics,
and statistical sciences. 
They are powerful rivals of MCMC 
in these fields. 

The aim of this paper is {\it neither\/} 
the design of a novel algorithm
{\it nor\/} the presentation of new applications. 
Our goal is to give a minimal set of references and explanations
for the cross-fertilization 
among the researchers in different fields. It will be
useful because few people realize essentially
the same algorithms are used in both in statistical information processing~\cite{K96,K98,LC98,book,Crisan00}
and physics~\cite{H84,CM,K62,CK,NB,G00}.
I also hope that this survey will be useful for the 
development of applications in machine learning 
and probabilistic artificial intelligence. 

In Sec.~\ref{S1}, we give a common structure of the
population Monte Carlo algorithms. 
We also discuss a relation to genetic
algorithms and give remarks on the origins of the
methods. In Sec.~\ref{S2}, we give
examples of the algorithms both in physics and
statistics. 

\section{An Overview}
\label{S1}
\subsection{Algorithm -- General}

Essentially, the
algorithms discussed in this paper 
are designed for the computation of the products 
of non-negative, sparse, $M \times M$ 
matrices $G^1, G^2, \ldots$ and a non-negative vector $X^{0}$.
If we express the vector of $n$th step as $X^{n}$ then 
\begin{equation}
\label{prod}
X^{n+1} (i) = \sum_j G^n (i,j) \, X^n (j)
\end{equation}
where $1 \leq i,j \leq M$ are indices of components
of the vector $X^n$ and the matrix $G^n$.
Here, the dimension $M$ of vector $X^{n}$ is assumed to be very large
and we cannot explicitly store the elements of the
vector in the memory. Saving the storage, 
we represent the non-negative vector as a weighted superposition
of the ``walker'' or ``particles'' indexed by $k$ $(k=1, \ldots, K)$, 
each of which is
placed at $j^n(k)$ with a weight $w^n_k$:  
\begin{equation}
\widetilde{X}^n (j) = \sum_k w^n_k \cdot \delta_{j^n(k),j}
\end{equation}
Here, $\delta$ is the Kronecker delta
defined by
\begin{eqnarray}
\delta_{m',m} & = 1 & (m'=m) \nonumber \\ 
              & = 0 & (m' \neq m) 
\end{eqnarray}
The algorithm is defined by the iteration of
the following steps:
\begin{enumerate}
\item {\sf STEP 1}: Random walk in the state space. \\
Each walker $k$ at the position $j^n(k)$ is moved 
independently to a new position $i$ 
\begin{equation}
j^{n+1}(k):=i
\end{equation}
according to the probability 
\begin{equation}
P^n (i,j^n(k)) = \frac{G^n (i,j^n(k))}
{\sum_{i'} G^n (i',j^n(k))} {}_,
\end{equation}

\item
{\sf STEP 2}: Update of the weights. \\
For each walker, calculate the factor
\begin{equation}
W_k = \sum_{i'} G^n (i',j(k))
\end{equation}
and update the weight of the walker using
\begin{equation}
\log w^{n+1}_k := \log W_k + \log w^n_k {}_. \label{weight}
\end{equation}

\item {\sf STEP 3}: Reconfiguration. 
\begin{itemize}
\item Split walkers with a large weight.
Each walker splits
into multiple walkers, whose total weight is
equal to that of the original walker. 
\item Remove walkers with a small weight.
\end{itemize}
This procedure is called by various different names
--- {\bf reweighting}, {\bf prune/enrichment}, {\bf reconfiguration}, 
{\bf rejuvenation}, {\bf resampling}, {\bf branching},
{\bf selection}.

There are several different ways to resample the
population of the walkers. A simple way is to set 
\begin{equation}
Q_k= \frac{w_k}{\sum_{k'} w_{k'}}
\end{equation}
and resample the walkers with probability $Q_k$
(Note that walkers with large $Q_k$ can be sampled
several times.).  Variants are found in the references, 
for example, \cite{NB}. In some variants, 
the number of walkers $K$ is not strictly 
constant, but fluctuates within a range.
\end{enumerate}

To complete the algorithm, the initial positions $\{j^0(k)\}$
and weights $\{w^0_k\}$ of the walkers should be given. 
In most cases, $w^0_k$ is a constant independent of $k$ and 
$\{j^0(k)\}$  are independent samples from the initial
density defined by $X^0$. The choice of $X^0$ 
depends on a specific problem and usually the same as
the choice of $X_0$ in the corresponding deterministic
algorithm based on the direct iteration of~eq.~(\ref{prod}).
For example, we can use an arbitrary density
as an initial vector $X_0$ in quantum Monte Carlo~(Sec.~\ref{Q}), 
although systematic choice will improve the speed of the 
convergence.
In Monte Carlo filtering~(Sec.~\ref{B1}), $X_0$ represents 
a prior distribution
for initial states of  series. 

Let us define $[ \cdots ]$ as an average over
random numbers used in the past steps of the algorithm.
Then, if we assume
\begin{equation}
X^n (j)= [ \widetilde{X}^n (j) ] {}_,
\end{equation}
it is not difficult to show that the relation
\begin{equation}
\label{basic}
X^{n+1} (j)= [ \widetilde{X}^{n+1} (j) ]
\end{equation}
holds with {\sf STEP 1} $\sim$ {\sf STEP3}.
This property is essential for the
adequacy of population Monte Carlo algorithm.
When we modify the {\sf STEP 3}, we should be
careful to conserve this property.
With the condition~eq.~(\ref{basic}),
we expect the convergence of the algorithm in the
limit of the number of the walkers $K \rightarrow \infty$ with
a fixed number $n$ of the iteration~\footnote{Note 
that the convergence is not
assured in the limit $n \rightarrow \infty$ 
with a fixed number $K$ of walkers. 
It is an important remark for the problems
where the limit $n \rightarrow \infty$ is required 
(for example, see~Sec.~\ref{Q}).}. 
We refer to~\cite{Crisan00} for a recent 
result on the convergence of population Monte Carlo 
algorithms~\footnote{Some remarks by physicists
on the property of the algorithm  
are found in~\cite{H84,CM}.}. 
Practically, the rate of the convergence
severely depends on the problem treated by the method.

Finally, we note that
it is easy to generalize the algorithm to the cases 
with continuous state space, where the matrix $G$
is replaced by an operator and the summations become integrals.
Such continuous versions of the algorithm are successfully used 
in some of the examples discussed below, {\it e.g.,} applications to 
quantum many-body problems.

\subsection{The Role of {\sf STEP 3}}

The {\sf STEP  3} of the algorithm is not necessary 
for the validity of the algorithm, i.e.,
the condition that we discussed in the previous
section is satisfied only with {\sf STEP 1} and {\sf STEP 2}. 
The role of {\sf STEP~3} is to suppress the variation of weights
$\{w_k\}$ and improve the efficiency of the algorithm.

For some problems, the algorithm without {\sf STEP  3} is
sufficient. Hereafter, we call the algorithm
without {\sf STEP  3} as ``simple'' population Monte Carlo. 
Note that without {\sf STEP  3}, there is no interaction 
between walkers. Then, parallel evolution of the
walkers can be replaced by sequential runs
each of which corresponds to the simulation of a walker.

In some references, a part of {\sf STEP 3} is 
separately discussed as a ``population control'' 
procedure ({\it e.g.,}~\cite{K62,PERM}) and/or
a part of {\sf STEP 3} is merged into 
{\sf STEP 1}~({\it e.g.},~\cite{K62}).
In such cases, we should be careful to judge whether the
algorithm effectively contains {\sf STEP 3} or not.

\subsection{Relation to Genetic Algorithm}

There is an obvious analogy between population Monte Carlo
and genetic algorithms (GA)~\cite{H97}. 
However, there is an essential difference in the goal
of the algorithm. Population Monte Carlo is a tool
for the computation of the product of matrices, 
multiple summations (integrals) and calculation of
marginals. On the other hand, GA is a tool specialized in
optimization. Usually, we can modify population 
Monte Carlo for the optimization. However, the converse
is often not true. Specifically, we should be
careful to the introduction of 
{\it crossover operators} and other tricks popular
in GA to population Monte Carlo,
because they can easily spoil 
the convergence to the exact result in the limit of 
infinite number of walkers, $K \rightarrow \infty$.

There are some theoretical studies on GA that assume 
the absence of the
crossover operation. It might be interesting to apply
them to the study of population Monte Carlo. 

\subsection{Origins}

It is difficult to fix the origin of the population Monte Carlo.
An origin seems the algorithms for the calculation
of the elements of an inverse Matrix developed by 
von Neumann and others~\cite{HH,JD}, which lacks
the {\sf STEP 3}. Another reference is 
Metropolis and Ulam~\cite{MU}~\footnote{
It is interesting to point out that this paper by famous physicists
appeared in the Journal of American Statistical Association, which
is one of the well-known journals of statistics.}, which discuss
a solver of the Schr\"{o}dinger equation by simulation. The algorithm is
attributed to Fermi (It is an origin of quantum Monte Carlo algorithms
with random walkers discussed in Sec.~\ref{Q}). The algorithm for
eigenvalue problems are also discussed in the classic 
book~\cite{HH} on Monte Carlo.
Simple algorithm for self-avoiding walks is introduced
by Rosenbluth and Rosenbluth~\cite{RR}. According to~\cite{Crisan00},
one of the earliest studies of simple algorithm 
in statistical science is found in~\cite{HM69}.

It is also not easy to determine who
introduced {\sf STEP~3}. For example, {\sf STEP~3} is included in
the algorithm in Metropolis and Ulam~\cite{MU}
and Kalos~\cite{K62} in a somewhat implicit manner.
The ``enrichment'' algorithm~\cite{WE} 
for self-avoiding walks contains
a very special case of {\sf STEP 3}. 
Anyway, many authors have used
the algorithms with {\sf STEP 3}  by the end of 1980s 
as a version of
``quantum Monte Carlo'' or 
``transfer-matrix Monte Carlo'', while the algorithms without
{\sf STEP  3} have also been used up to now. 

We left a comprehensive treatment on the
history of the population Monte Carlo methods 
to future surveys.

\section{Examples}
\label{S2}
\subsection{Quantum Many-Body Problems}
\label{Q}

First, we discuss ``quantum Monte Carlo'' 
algorithms~\footnote{There is a different type of ``quantum Monte Carlo'' algorithms
which are based on MCMC (dynamical Monte Carlo). They simulate a ``world-line'' instead of a collection of random walkers.}~\cite{CK,SK,SC,DL,H84,CM,N86,A75,BS,JR}.
In spite of their name, they are not algorithms specialized to
quantum mechanics. 
They are most naturally understood as
methods for the approximation of the smallest (or largest) eigenvalue 
and the corresponding eigenvector of
a large sparse matrix. Essentially, they are stochastic versions 
of the power method for eigenvalue problems.

Consider a symmetric, non-negative 
matrix $A$. To formulate the eigenvalue 
problem into population Monte Carlo, we introduce the exponential
$\exp(-\beta A)$ of $A$. For sufficiently large $\beta$, the expression
\begin{equation}
\exp(-\beta A) \, X^0
\end{equation}
approximately gives  an eigenvector with the smallest eigenvalue for 
an arbitrary vector $X^0$ that has non-zero projection to 
the eigenvector. 
In general, the calculation of $\exp(-\beta A)$ is
difficult. In most of interesting problems, however,
 $A$ is decomposed into several simple components, {\it e.g.,}
$A= B+C$ where $B$ and $C$ is symmetric, non-negative matrices whose
exponential is easily calculated. 
Thus, the exponential of $A$ is expressed as
\begin{equation}
\hspace{-23.pt}
\exp(-\beta A) = \lim_{N \rightarrow \infty}
\left ( \exp(-\frac{\beta}{N} B) \exp(-\frac{\beta}{N} C) \right )^N.
\end{equation}
It is straightforward to fit the problem 
into the form of the equation~eq.~(\ref{prod}) with
\begin{equation}
G^{n} =\exp(-\frac{\beta}{N} B) \exp(-\frac{\beta}{N} C)
\end{equation}
where $N$ is a sufficiently large number and $n=0, \ldots, N-1$.

Another approach to the eigenvalue problem is 
a method based on an
iterative solver of the eigenvalue equation.
Consider the product of matrices
\begin{equation}
G^{n}= B^{-1} (\lambda I-D)
\end{equation}
where $A=B+D$ and $D$ are a diagonal matrix
($I$ denotes an identity matrix.). Under the suitable condition
on the ``Green's function'' $B^{-1}$ and a properly chosen value
of $\lambda$~\footnote{When we implement it in a form of population 
Monte Carlo, the choice of the constant $\lambda$ is
absorbed in {\sf STEP 2} and {\sf STEP 3}. Then we can obtain
the eigenvalue $\lambda$ and the corresponding 
eigenvector simultaneously.}, 
the vector $X^n$ converges to an eigenvector 
of $A$.

In the treatment on 
quantum mechanical problems, the matrix $A$ is usually {\it Hamiltonian\/}
of the given system. It is easy to construct continuous version
of the algorithm, where the matrices and summation is replaced by
operators and integrals, respectively. 
In such cases, the operator $B$ is often 
represented by the summation of Laplace operators and a constant.

These algorithms are successfully applied to various problems in
quantum physics, {\it e.g.,} the computation of the properties of
nuclei~\cite{K62,N86}, 
superfluids~\cite{CK,SC}, 
quantum spin systems~\cite{BS,DL}, 
and quantum dynamics~\cite{JR}. 
They are called  by the terms ``projector Monte Carlo'', 
``Green's function Monte Carlo'', and ``diffusion Monte Carlo''~\footnote{
Here we avoid the details of the terminology, because the 
classification and naming of these algorithms may depend
on the authors.}.

In some cases, we should deal with matrices or eigenvectors 
that are not non-negative. Formally, these cases can be dealt with
the introduction of negative weight $w^n_k$ of walkers.
It, however, seriously spoils the efficiency and convergence
of the algorithm. It is an example of ``negative sign crisis''
in stochastic computation of quantum mechanical problems. 
Although there are some
interesting ideas to improve the algorithm~\cite{K84,CaK,NeB,AT,LZK}, 
it seems difficult to remove the difficulty entirely with a smart
trick, because
it originates from radical difference 
between classical stochastic systems and
quantum systems. 

\subsection{Lattice Spin Systems}
\label{L}

Using the transfer matrix (transfer integral) formalism, we can
translate a calculation for classical statistical mechanics (and combinatorics)
to the computation of a product of matrices. Then,
it is a natural idea to apply population Monte Carlo algorithms
to these problems. For example, consider a 
classical spin model (Markov field model)
defined on  a $L_1 \times L_2 \, (L_2>>L_1)$ strip~\footnote{For a
three-dimensional problem, a 
$L_1 \times L_1 \times L_2$ column
substitutes for the strip.}. We define a set of subsystems
each of which consists of the $1, \ldots, n$ rows of the original
strip ($n=1,2, 
\ldots, L_2$). 
Hereafter we assume nearest neighbor interaction
on the lattice, although 
extensions to cases with next-nearest interaction {\it etc.\/} 
are formally easy.
The partition function $Z_n(S_n)$ of 
the subsystem $n$ conditioned with the values 
of the variables in the $n$th row
$S_n= \{S_{ln}\}_{l=1 \ldots L_1}$  is
\begin{equation}
\sum_{S_1} \sum_{S_2} \cdots \sum_{S_{n-1}} 
\exp(-\beta E_n(S_1,S_2, \ldots, S_n))
\end{equation}
where $\beta$ is inverse temperature and $E_n(S_1,S_2, \ldots, S_n)$
is the energy of the $n$th subsystem.
When each variable $S_{ln}$
is a binary variable, the number of the possible values of 
$S_n$ is $2^{L_1}$.
Then, a recursion relation
\begin{eqnarray}
\lefteqn{Z_{n+1}(S_{n+1}) =} & \nonumber \\
 & \sum_{S_n} Z_n(S_n) \exp(-\beta \triangle E_n(S_n,S_{n+1}))
\end{eqnarray}
where
\begin{eqnarray}
\lefteqn{\triangle E_n(S_n,S_{n+1}) =} & \nonumber \\
& E_{n+1}(S_1, \ldots, S_n, S_{n+1}) - E_n(S_1, \ldots, S_n)
\end{eqnarray}
holds. 
Obviously, it is essentially the same as
the filtering formula for general state space models and 
the recursive formula of the Baum-Welch algorithm for
hidden Markov models. 
We define $Z_0=1$ and $\triangle E_0(S_1)=E_1(S_1)$
for the free boundary condition at the top of the strip
(Here and hereafter, 
the symbol $S_0$ in the formulae should be neglected.).

A population Monte Carlo algorithm
is constructed by setting $G$ and $X$ as
\begin{equation}
G^n(S_{n+1},S_n) = \exp(-\beta \triangle E_n(S_n,S_{n+1}))
\end{equation}
\begin{equation}
X^n(S_n) = Z_{n}(S_n) {}_.
\end{equation}
The algorithm is called ``transfer-matrix Monte Carlo''
and applied to classical Heisenberg and XY
models~\cite{NB,TK} and enumeration of the number of Penrose tiling~\cite{SH}
on a lattice.

\subsection{Bayesian Computation: MCF and Sequential Monte Carlo}
\label{B1}

Algorithms that are similar to the ``transfer-matrix Monte Carlo''
are developed by statisticians for the models of time-series analysis
and models with sequential structures ({\it e.g.,} models for
gene-propagation analysis). In these algorithms, we consider
marginal likelihood conditioned with the state $x_n$ of the latest
step (time-step) $n$
\begin{eqnarray}
\lefteqn{Z_n(x_{n}) = \sum_{x_0} \sum_{x_1} 
\cdots \sum_{x_{n-1}} \hat{P}(x_0) \tilde{P}(x_0|x_1) 
} & \nonumber \\ 
& \hspace{-0.pt}
P(y_1|x_1) \cdots \tilde{P}(x_{n-1}|x_n) P(y_n|x_n) \nonumber \\
\end{eqnarray}
instead of partition function~\footnote{
We can also use formulation based on posterior
distributions of subsystems, which is similar to 
the one used in~Sec.~\ref{P}. It is useful in some
problems, {\it e.g.,} smoothing of time-series.}. 
Here
$P(y|x)$ is likelihood associated 
with data $y$ (representation of observational noise). 
$\tilde{P}(x|x')$ and $\hat{P}(x)$ are prior probability
for the state transition $x \rightarrow x'$ 
(representation of system noise) and 
prior probability of initial state $x$, respectively. 
The application of population Monte Carlo algorithms 
is straightforward.

Kitagawa~\cite{K96,K98} introduced
``Monte Carlo filter'' (MCF, ``particle filter'')
for non-Gaussian/non-linear state space models. 
Gordon and coworkers independently developed an idea
similar to the one by Kitagawa, which 
they call ``bootstrap filter''~\cite{GSS,GSE}.    
Another contribution is 
``sequential imputation'' that first used 
in gene-propagation analysis~\cite{ICK} and later used for
blind deconvolution~\cite{LC95} and other problems. The original
form of sequential imputation is a simple algorithm
that lacks {\sf STEP 3}, but {\sf STEP 3} is introduced later
(``rejuvenated sequential imputation'').
The term ``sequential Monte Carlo'' is also used
to indicate a family of 
algorithms that include the above-mentioned methods 
as special cases~\cite{LC98,Crisan00}. In \cite{LC98,Crisan00}, 
more references in statistical sciences, some of which are earlier
than above-mentioned studies, are discussed.
Applications in robot vision are described in~\cite{I98}.
A forthcoming book~\cite{book} on sequential Monte Carlo 
by statisticians will be a good guide 
to the applications in statistical sciences.

\subsection{Polymer Science}
\label{P}

``Polymer'' means a flexible chain consists of 
small molecules (``monomers'') connected in a fixed order.
It is called a heteropolymer ($\leftrightarrow$ homopolymer)
when it is made by monomers 
of more than two species. Well-known examples of 
heteropolymer is protein, RNA and DNA.
A good model of polymer is a flexible chain
where each unit is mutually repulsed by short range
force (van der Waals core) and repulsed/attracted by 
longer range force. In lattice models, the former interaction
is modeled by ``self-avoiding condition'', i.e., the condition that
a lattice point should not be occupied  by more than two monomers.

Population Monte Carlo algorithms for the study of 
polymer models are designed in a manner similar to
those for lattice spin models. Subchains of
length $n=1,2, \ldots, N$ that consists of the first $n$ monomers of 
the original polymer of length $N$ is used 
as subsystems for the definition of the
algorithm. An important difference is that the index of a vector
$X^n$ is {\it not\/} the coordinates $r_n$ of the 
monomer located at the end of a subchain, 
but coordinates $\{r_n'\}, (n'=1, \ldots, n)$ of all monomers
in the subchains. This is because a pair of monomers with 
a large distance along the chain can interact each other, even
with short range interaction.
Thus, the recursion is better described by
the Gibbs distribution $P_n$ of subchains, rather than partition
functions $Z_n$. 

If we express the Gibbs distribution of a subchain $n$ as
$P_n(r_1, r_2, \ldots, r_n)$,
the recursion relation   
\begin{eqnarray}
\label{R}
\lefteqn{P_{n+1}(r_1, r_2, \ldots, r_{n+1}) \propto} &  \nonumber \\ 
& \hspace{-20.0pt}
P_n(r_1, r_2, \ldots, r_{n}) \cdot
\exp(-\beta \triangle E(\{r_1,\ldots,r_{n}\}, r_{n+1})) \nonumber \\ &
\end{eqnarray}
holds, where $\triangle E(\{r_1, \ldots, r_{n}\}, r_{n+1})$
is the interaction energy between 
a new monomer and the monomers
in the subchain $n$. For lattice models, 
we set $\triangle E=+\infty$ 
if the conformation of the subchain specified 
by $\{r_1, \ldots, r_{n+1}\}$ is non-physical, i.e., 
self-avoiding condition or 
connectivity condition of the chain is violated. 

To define a population Monte Carlo algorithm,
we must specify the index $i$ of $X^n(i)$ in the equation~eq.~(\ref{prod}).
The index $i$ should determine the conformation 
$\{r^i_1, \ldots, r^i_{n} \}$
of the $n$th subchain and should have
the dimension $N$ independent of $n$. 
Here we define $i$ as a vector with components
$i_s=r^i_s (s \leq n)$
and $i_s={\bf 1} (s > n)$, i.e.,
\begin{equation}
\label{iii}
i = (r^i_1, r^i_2, \ldots, r^i_n, {\bf 1}, {\bf 1}, \ldots, {\bf 1} )
\end{equation}
where ${\bf 1}$ is the coordinates of an arbitrary point,
for example, the origin.  
Next we define $X^n(i)$ as 
\begin{equation}
X^n(i) \propto P_n(r^i_1, r^i_2, \ldots, r^i_{n})
\end{equation}
for $i$ of the form~eq.~(\ref{iii}) and set $X^n(i)=0$
for other indices $i$.
Finally, we define the matrix $G^n$ as
\begin{eqnarray}
\lefteqn{G^n_{ij} = 
\exp(-\beta \triangle E(\{r^i_1, \ldots, r^i_{n}\}, r^i_{n+1}))} & \nonumber \\
& \times  \prod^n_{s=1} \delta (i_s - j_s) \, 
\prod^{N}_{s=n+2} \delta (i_s - {\bf 1}) {}_.
\end{eqnarray}
With these expressions,
we can write the recursion~eq.~(\ref{R}) in the form of
the equation~eq.~(\ref{prod}), which 
gives population Monte Carlo algorithms
based on a simulation of growth/selection of polymer subchains
(here we start from $X^1$ instead of $X^0$, with which
the index $n$ equals to the length of the subchain).

Population Monte Carlo algorithms
for polymer models are developed by two or more
groups (\cite{GO,O98} and \cite{G97,PERM,BFGGN,G00})~\footnote{ 
A related problem (counting of ``meanders'') is
treated in~\cite{Gol00} by population Monte Carlo.}.
Grassberger pointed out that his algorithm,
PERM (pruned-enriched Rosenbluth method) is understood
as an unification of ``enrichment algorithm''~\cite{WE}
 ({\sf STEP 1} and {\sf 3}) and
``Rosenbluth algorithm''~\cite{RR,Seno} 
({\sf STEP 1} and {\sf 2}). PERM has proved
to be an useful tool to study 
statistical mechanics of long homopolymers~\cite{G97} and
other lattice statistical models~\cite{PERM,G00}.
It is also the most efficient
algorithm to compute finite temperature properties of
lattice protein models~\cite{BFGGN,G00}, although it is recently challenged 
by a MCMC algorithm developed by the author and coworkers~\cite{CKI}. 

\subsection{Bayesian Computation: Annealed Importance Sampling}
\label{B2}

In the example of Sec.~\ref{Q}, an intermediate state indexed by $n$ is
a fictitious one and we are only interested in the limit
of large $n$. 
In Sec.~\ref{L}--\ref{P}, we dealt with examples
where the index $n$ corresponds to an index in the real world, {\it e.g.,}
the index of row of a lattice, time-step of series, 
and the length of the subchain.
Here we consider cases where the index $n$ indicates a system with
a parameter $\gamma_n$,
{\it e.g.,} temperature in statistical physics
and hyperparameters in Bayesian models. Such an example is discussed
by Neal and called by ``annealed importance sampling''~\cite{N98}.

Consider a parametric family of distributions $\{P_{\gamma_n}(x),
n=1, \ldots, N\}$.
An example is
\begin{equation}
P_{\gamma_n}(x) = \{P(y|x)\}^{\gamma_n} \tilde{P}(x)
\end{equation}
where $P(y|x)$ is likelihood with data $y$ and
$\tilde{P}(x)$ is a prior. We can formally define
a ``recursion relation'' by
\begin{equation}
\label{rew}
P_{\gamma_{n+1}} (x)= \frac{P_{\gamma_{n+1}}(x)} 
{P_{\gamma_{n}}(x)}
\cdot P_{\gamma_{n}}(x) {}_.
\end{equation}
Unfortunately, population Monte Carlo algorithm
derived from~eq.~(\ref{rew}) is trivial and inefficient, because 
the matrix $G^n(x,x')$ defined by the recursion is diagonal.

An idea by Neal is to introduce MCMC steps into the
algorithm.  The insertion of probabilistic 
state-changes that makes
$P_{\gamma_n}(x)$ invariant corresponds to the insertion
of a non-diagonal matrix $\widehat{G}^n$ to the right-hand-side of 
the equation~eq.~(\ref{prod}), which satisfies
\begin{equation}
X^{n}(i)=\sum_j \widehat{G}^n(i,j) \, X^n(j) {}_.
\end{equation}
Obviously, the introduction of such $\widehat{G}^n$ does not
have an effect on the validity of the results.
The adequacy of the procedure is also ensured by
the relation
\begin{equation}
X^n (j)= [ \widetilde{X}^n (j) ]
\end{equation}
is not changed by the addition of MCMC steps.
The resultant algorithm consists of
{\it finite\/} MCMC sweeps at each $n$ 
and {\sf STEP 2} with
\begin{equation}
W_k = \frac{P_{\gamma_{n+1}}(x)}{P_{\gamma_{n}}(x)}
\end{equation}
where $x=j(k)$ is the position of the $k$th walker.

As far as I know, Neal does not discuss algorithms with 
{\sf STEP 3}, but there seems no reason against 
the use of {\sf STEP 3} in the algorithm.
On the other hand, as Neal himself pointed out,
the algorithm without {\sf STEP 3} can be regarded as a version
of non-adiabatic thermodynamic integration discussed 
by Jarzynski~\cite{J97,H01}.

It is interesting to remark that the introduction of 
MCMC steps~\cite{Crisan00}
is possible for some of 
the other population Monte Carlo 
algorithms.

\section*{Acknowledgments}
This paper is a revised version of the paper presented
at 2000 Workshop on Information-Based Induction Sciences
(IBIS2000, Syuzenji, Japan, 2000, July).
We would like to 
thank to Prof.~T.~Higuchi who kindly taught us about the 
book~\cite{book} and some of other references in~Sec.~\ref{B1}. 
We also acknowledge Dr.~A.~Doucet for giving us kind advices
and significant information on references.

\end{document}